\documentclass[manuscript]{acmart}
\usepackage{tabularx}

\AtBeginDocument{%
  }

\setcopyright{acmlicensed}
\copyrightyear{2026}
\acmYear{2026}
\acmDOI{XXXXXXX.XXXXXXX}

\acmConference[Conference acronym 'XX]{Make sure to enter the correct
  conference title from your rights confirmation email}{June 03--05,
  2018}{Woodstock, NY}
\acmISBN{978-1-4503-XXXX-X/18/06}

\begin{document}

\title[Transparency Practices in Multi-Agent LLM Systems]{``So There's a Catch-22 Here'': How Early Adopters Who Build Multi-Agent LLM Systems Conceptualize Transparency}


\author{Suchismita Naik}
\authornote{This work was done during internship at Microsoft Research}
\email{naik33@purdue.edu}
\orcid{0009-0002-5667-4576}
\affiliation{
    \institution{Purdue University}
    \city{West Lafayette}
    \state{Indiana}
    \country{USA}
}

\author{Samir Passi}
\email{v-sapassi@microsoft.com}
\orcid{0000-0002-7921-3820}
\affiliation{%
  \institution{Cornell University}
  \city{Ithaca}
  \state{New York}
  \country{USA}
}

\author{Mihaela Vorvoreanu}
\email{Mihaela.Vorvoreanu@microsoft.com}
\orcid{0000-0002-3322-3548}
\affiliation{
    \institution{Microsoft Research}
    \city{Redmond}
    \state{WA}
    \country{USA}
}

\author{Scott Saponas}
\email{ssaponas@microsoft.com}
\orcid{0000-0002-8806-0125}
\affiliation{
    \institution{Microsoft Research}
    \city{Redmond}
    \state{WA}
    \country{USA}
}

\author{Amanda K. Hall}
\email{amanda.hall@microsoft.com}
\orcid{0000-0001-6151-1814}
\affiliation{
    \institution{Microsoft Research}
    \city{Redmond}
    \state{WA}
    \country{USA}
}

\renewcommand{\shortauthors}{Naik et al.}

\begin{abstract}
Multi-agent large language model (LLM) systems are rapidly emerging, yet transparency, a cornerstone of responsible AI, remains under-defined in these distributed architectures, which have complexities of inter-agent coordination and orchestration. In this paper, we present one of the first empirical study of how early adopters of multi-agent LLM systems, who are both the builders and users, understand and practice transparency. We conducted semi-structured interviews with 13 early adopters in [Large Technology Organization] and applied thematic analysis to identify recurring patterns. Participants articulated divergent yet complementary framings of transparency, including reproducibility, debugging, boundary-setting, visualization, and auditing. These perspectives spanned questions of what transparency entails, why it matters, and how it is achieved. We synthesize these into a multidimensional framework, which is developer, user, and governance-focused positioning transparency as a situated socio-technical practice that informs future HCI and AI design and research around aligning expectations and capacities of their intended audiences.

\end{abstract}

\maketitle

\section{Introduction}
Multi-agent generative AI (GenAI) systems are evolving rapidly, with large language models (LLMs) being increasingly deployed in orchestrated configurations in which multiple agents collaborate to complete complex tasks. These systems are beginning to shape workflows in domains ranging from software development and product design to business operations and creative industries \cite{naik_exploring_2025}. Their promise lies in enabling new forms of automation, coordination, and problem-solving. However, alongside this promise comes a pressing challenge: what transparency should entail, how it should be defined, achieved, and sustained when intelligence is distributed across interacting agents rather than confined to a single model.

Transparency is now widely regarded as integral to good AI design and a cornerstone of responsible AI (RAI), underpinning accountability, trust, and ethical governance. However, its meaning and implementation remain fragmented and unclear. Existing frameworks, such as explainable AI (XAI) techniques, model cards, or interpretability metrics, were largely developed for non-agentic systems. Agentic architectures introduce new forms of opacity, as task delegation, inter-agent communication, and orchestration mechanisms complicate traditional notions of transparency. These dynamics create uncertainty not only for developers who must debug and refine such systems, but also for users who rely on them and institutions that seek to regulate them. This complexity motivates our central research question: \textit{How do early adopters of multi-agent generative AI systems perceive transparency, and what implications does this have for designing transparent systems?}

Early adopters occupy a unique position in addressing this question. While not all of them may be experts in AI ethics or transparency, they interact both as developer and user, with agentic systems during their formative stages, when norms and practices are still unsettled. Their perceptions matter because they shape how transparency is enacted in practice, through design choices, debugging strategies, documentation, and everyday usage. Studying these early practices allows us to surface challenges and opportunities that are likely to influence the future trajectory of multi-agent AI adoption and design. In this paper, we make four key contributions:
\begin{enumerate}
    \item We provide one of the first empirical studies of how early adopters understand transparency in multi-agent LLM-based systems, offering grounded insights into an emerging domain.
    \item We identify divergent yet complementary framings of transparency, including reproducibility, boundary-setting, debugging, visualization, trust-building, compliance, and auditing, and highlight how these perceptions point to transparency as a ``situated practice.''
    \item We synthesize these perspectives, taking initial steps towards a multidimensional framework of transparency that highlights tensions and synergies across developer-, end-user-, and governance-focused transparency needs.
    \item We discuss implications for HCI and AI design, emphasizing transparency's socio-technical aspects that must be tailored across stakeholder groups and contexts of use.
\end{enumerate}

In the following sections, we first review related work on AI transparency, multi-agent systems, and early adopters as key stakeholders. We then describe our methodology, including participant recruitment, data collection, and analysis. The findings present early adopters' diverse interpretations of transparency, while the discussion synthesizes these into a multidimensional framework and explores the implications for design and governance. We conclude by reflecting on the significance of studying transparency in the early stages of multi-agent AI adoption and design.

\section{Related Work}

\subsection{Transparency in AI Systems}
The growing integration of AI systems into diverse sectors highlights the paramount importance of transparency for effective human-AI interaction and collaboration and broader societal acceptance \cite{caldwell_agile_2022, vossing_designing_2022, eigner_determinants_2024}. Without transparency, AI systems frequently operate as ``black boxes,'' making their reasoning and decision-making processes obscure to users \cite{ngo_exploring_2020, eslami_user_2019}. This inherent opacity can lead to user discomfort, skepticism, or a lack of trust, which can impede AI adoption and collaborative outcomes \cite{adadi_peeking_2018, ngo_exploring_2020}. Trust is a fundamental element in effective teamwork, fostering confident collaboration and coordination, and in human-AI contexts, it is a complex social construct influenced by system properties, user perceptions, and environmental factors \cite{wang_investigating_2024, zhang_i_2024}. Consequently, designing for transparency is crucial to ensure accountability, facilitate calibrated trust, and ultimately enhance the decision-making and performance of human-AI teams. The increasing complexity and autonomy of AI systems have often led to a decrease in critical decision-making, further amplifying the need for transparency \cite{passi_appropriate_2024}.

To address AI opacity, the field of \textbf{XAI} has rapidly expanded, focusing on making AI outputs more comprehensible to humans \cite{ehsan_human-centered_2022, adadi_peeking_2018}. XAI aims to achieve three key desiderata: improve users' understanding of the AI model, help them recognize model uncertainty, and support their calibrated trust in the model \cite{wang_are_2021}. Explanations can be categorized as global, describing the overall model behavior, or local, focusing on the reasons for specific predictions \cite{wang_are_2021, adadi_peeking_2018}. Common XAI techniques include \textit{feature importance and contribution explanations}, which illustrate how features influence a prediction, and \textit{example-based or counterfactual explanations}, which provide illustrative scenarios to make reasoning more understandable to humans \cite{wang_are_2021}. Users' understanding can be assessed through tasks such as ranking feature influence, predicting model changes, simulating predictions, asking ``what-if'' questions, and detecting/debugging errors \cite{wang_are_2021}. However, the effectiveness of XAI can be inconsistent, particularly when users have limited domain expertise, as explanations may be difficult to absorb or to infer insights from \cite{morrison_impact_2024}. Additionally, explanations, especially if imperfect or incorrect, can backfire, leading to increased over-reliance, user overconfidence, or even deception \cite{li_assessing_2023, morrison_impact_2024, passi_appropriate_2024}. 

Beyond algorithmic explanations, standardized documentation tools such as model cards \cite{liao_designerly_2023, naik_designing_2025}, data sheets \cite{liao_designerly_2023, wang_are_2021}, AI service fact sheets \cite{caldwell_agile_2022}, guidelines for human-AI interaction \cite{amershi_guidelines_2019}, and human-AI experience toolkit \cite{vorvoreanu_create_2023} have emerged to promote the transparent reporting of AI model capabilities, limitations, and intended use cases, thereby supporting \textbf{Responsible AI (RAI)} practices and governance. These tools are intended to help practitioners and end-users evaluate model suitability and facilitate accountability. \textit{Traceability}, which links product requirements to system design, is another critical concept, especially in complex sociotechnical AI systems, where it can be difficult to trace model output to requirements \cite{raji_closing_2020}. User experience designers are seen as having a pivotal role in RAI practices by bridging user needs and technical affordances, anticipating harms, and developing mitigation strategies \cite{liao_designerly_2023}.

\subsection{Multi-agent Generative AI Systems and Challenges to Transparency}
The advent of multi-agent GenAI systems (or multi-agent LLM-based AI), collaborative systems comprising multiple LLM-based agents that dynamically reason, communicate, and adapt to achieve goals, introduces new and amplified transparency challenges \cite{naik_exploring_2025, banerjee_introduction_2022}. These systems leverage distributed problem solving and emergent intelligence, offering advanced capabilities for complex tasks. These GenAI systems are increasingly recognized as transformative technologies that augment human creativity, enhance processes, and enable novel forms of collaboration \cite{li_assessing_2023, weisz_design_2024}. However, their intricate architecture poses several hurdles for transparency.

The \textit{increasing sophisticated workflows} of multi-agent systems and \textit{inter-dependencies among agents} make it significantly harder to track decision-making processes and ensure predictable outcomes. Managing this complexity is a core design strategy \cite{naik_designing_2025}. One of the most pressing challenges is the \textit{lack of traceability}. It becomes challenging to trace agent-to-agent interactions, deepening the ``black box'' effect and hindering the ability to understand how agents influence each other, identify error sources, or debug effectively \cite{bansal_challenges_2024}. This necessitates attention to orchestration, traceability, and user oversight, which are not typically required in single-agent systems \cite{naik_designing_2025}. The probabilistic nature of AI further compounds these issues by introducing \textit{unpredictability and emergent behaviors} \cite{bansal_challenges_2024}. Errors from one agent can propagate, leading to compounded inaccuracies or ``unproductive loops'' where agents continuously interact without progress. In addition, \textit{human-agent communication} presents unique challenges. Establishing common ground between human users and AI agents is often ``fraught with error,'' as humans may struggle to interpret the complex dynamics of inter-agent interactions \cite{caldwell_agile_2022}. Interfaces must be designed to effectively convey these complex inter-agent dynamics while minimizing the cognitive load on humans \cite{bansal_challenges_2024}. The use of advanced AI in \textit{distributed decision-making} is a relatively nascent area of research, with much yet unknown about its mechanisms in spatially distributed hybrid teams \cite{wang_are_2021}. Shared understanding requires each team member (human \textit{and} AI) to grasp their roles, limits, and tasks \cite{caldwell_agile_2022}. Finally, an enduring challenge lies in \textit{balancing agent autonomy} with the necessary human oversight and control. While autonomous agents can contribute to efficiency and scalability, human intervention remains critical at decisive junctures \cite{naik_designing_2025, naik_exploring_2025}. Ensuring that oversight mechanisms are well integrated allows humans to intervene effectively without undermining the benefits of autonomy.

\subsection{Early Adopters of Multi-agent AI System}
Early adopters, including developers and creators actively engaged in designing, testing, and deploying multi-agent LLM-based AI systems, play a critical role in uncovering the practical challenges associated with these technologies \cite{naik_exploring_2025, choe_understanding_2014}. Acting as ``frontline co-designer[s]'' of emerging systems, their experiences generate valuable insights into usability limitations, transparency requirements, and the extent to which these systems align with or diverge from human expectations and mental models \cite{kulesza_tell_2012}. The feedback provided by these early users is instrumental in narrowing the gap between technical capabilities and user needs, thereby supporting the development of flexible, reusable, and human-centered applications \cite{naik_designing_2025}.

While existing studies extensively address transparency in single-system AI contexts and explore abstract goals related to fairness, explainability, and trust, there is a significant amount of research that is underexplored in understanding and supporting transparency specifically within multi-agent LLM-based AI systems \cite{liao_designerly_2023, zhang_i_2024, wang_are_2021}. Much of the existing research focuses on enhancing the capabilities of individual AI agents as isolated entities. However, multi-agent settings demand new analytical lenses that account for distributed actions, emergent behaviors, complex inter-agent dynamics, and the unique challenges of human-agent communication \cite{naik_designing_2025, bansal_challenges_2024}. Failures in these systems often originate not from individual agent performance, but from the intricate ways agents interact, delegate tasks, and interpret each other's outputs.

Research must therefore move beyond evaluating individual agent performance to focus on the comprehensibility of the collective system. This includes investigating how to design for orchestration-level transparency, which maps agent relationships and provides insights into their overall coordination. Furthermore, there is a need to understand transparency mechanisms that can cater to varying user expertise levels from technical to non-technical roles, offering simplified summaries for general users and detailed logs for experts, to ensure effective human sense-making and calibrated trust in these inherently more complex AI partnerships \cite{bansal_beyond_2019}. The ``human's role is not sufficiently studied in existing explainability approaches,'' and research in HCI and human sciences has often been isolated from XAI, highlighting the need for interdisciplinary collaboration to drive human-centric explainable models \cite{adadi_peeking_2018}. This study aims to contribute by exploring the transparency practices of early adopters, providing empirical insights to inform the human-centered design of future multi-agent LLM-based AI systems.

\section{Methodology}
To investigate how early adopters conceptualize and practice transparency in multi-agent LLM-based AI systems, we conducted semi-structured interviews with 13 participants. This sample size reflects the technology's nascent stage: multi-agent LLM systems remain experimental, with adoption concentrated among researchers and developers in major technology organizations. We therefore designed this study as foundational research, focusing on early adopters within a single organization to capture in-progress practices that remain largely inaccessible due to the novelty and limited diffusion of the technology.

We adopted an interpretivist approach \cite{creswell_qualitative_2018, soden_evaluating_2024} that allowed us to examine how transparency manifests in practice without constraining participant accounts to a predefined framework. Instead, we drew from multiple strands of HCI, AI, and transparency literature to sensitize our inquiry, while generating themes inductively from the data.

\subsection{Data Collection}
We conducted interviews that combined retrospective tasks and reflective activities with the participants. These activities prompted participants to describe prior experiences with developing or designing multi-agent LLM systems, to reflect on their everyday uses, and to articulate their interpretations of transparency. The interviews explored participants' motivations for adopting multi-agent LLM-based AI systems, their approaches to transparency, and their perceptions of the tools' capabilities and limitations. Drawing on Hoffman et al.'s \cite{hoffman_metrics_2018} notion of explicit explanation, we also asked participants to articulate how they understood and described multi-agent AI in their own words. Finally, we invited them to share applied examples of how they used these systems in everyday tasks, while ensuring no confidential project details were disclosed. 

Although we explicitly asked about transparency in one question, the concept continuously surfaced unprompted throughout participants' responses. Even when discussing motivations, limitations, or everyday workflows, participants repeatedly invoked transparency as a critical issue. This pervasiveness suggests that transparency was not only a topical prompt but also a foundational concern that participants considered integral to their engagement with multi-agent systems.

\subsection{Recruitment and Participants}
We recruited participants through three channels: (1) a screener survey distributed to individuals from a prior pilot survey shared in online internal developer communities; (2) recruitment emails circulated within [Large Technology Organization]; and (3) snowball sampling from these groups.

Eligibility criteria required participants to have first-hand experience developing or experimenting with multi-agent AI frameworks in a professional context, to possess at least intermediate knowledge of generative AI, and to be 18 years or older. We excluded senior executives to avoid perspectives shaped primarily by high-level managerial decision-making. Eligible participants completed a survey to confirm criteria and review informed consent. Participants who consented were scheduled for a 60-minute interview via Microsoft Teams. The study was reviewed and approved by [Large Technology Organization]'s Institutional Review Board.

The final cohort comprised 13 participants: five in non-technical roles (e.g., product managers and designers) and eight in technical roles (e.g., software developers and data scientists). All were employed at [Large Technology Organization] and were based primarily in the United States and Europe. Each participant reported developing either a proof-of-concept or a functional multi-agent LLM tool, which was actively used in daily work tasks such as automation, collaboration, and creative exploration. They employed diverse multi-agent AI frameworks, including AutoGen \cite{wu_autogen_2023, dibia_multi-agent_2023}, TaskWeaver \cite{qiao_taskweaver_2023}, BizChat, LlamaIndex, and custom-built systems. Participants cited motivations such as simplifying workflows, experimenting with system capabilities and limitations, and sharing knowledge with their colleagues.

\subsection{Data Analysis}
We recorded and transcribed nine hours of interviews using an internal transcription tool and then manually refined the transcripts. We organized responses by interview question and collaboratively conducted thematic analysis \cite{braun_thematic_2012} to surface recurrent themes and categorize them across participants' accounts. Through iterative coding, we generated 89 codes that captured participants' interpretations of transparency, as well as their views on the capabilities, limitations, and design implications of multi-agent AI.

Notably, transparency emerged as a pervasive theme, even in areas where we had not directly prompted it. This underscored its salience as a cross-cutting issue: participants treated transparency as inseparable from questions of adoption, system boundaries, debugging, and governance. Recognizing this, we refined our coding scheme to capture both explicit mentions of transparency and its implicit presence in adjacent practices.

Three researchers engaged in iterative discussions over a seven-week period to refine the coding scheme and ensure analytical rigor. This inductive process enabled us to derive themes that highlight how early adopters framed transparency as layered, multidimensional, and contingent on role, context, and practice. The subsequent findings section elaborates on these themes.

\section{Findings: Divergent Perceptions of Transparency in Multi-Agent LLM Systems}
Our analysis revealed that transparency is not a singularly defined concept among early adopters of multi-agent LLM-based systems. Instead, participants articulated divergent and often complementary understandings that spanned multiple dimensions: \textbf{what} transparency entails (e.g., internal visibility and scope of capabilities), \textbf{how} it is achieved (e.g., visualization interface, evaluation metrics, and traceability), and \textbf{why} it matters (e.g., open-source reproducibility, trust-building, and auditing mechanism) highlighted in Tables ~\ref{tab:transparency_findings1} and ~\ref{tab:transparency_findings2}. Their perspectives highlight that transparency operates simultaneously as a technical, ethical, and experiential construct. These findings highlight that transparency is a layered construct, often shaped by participants' roles, goals, and expectations in the design and deployment of such systems.

\subsection{The ``What'' of Transparency: Making Systems Understandable}

\subsubsection{Transparency clarifies Capability Boundaries and Limitations}

Several participants pointed to the need to make the boundaries of AI capabilities explicit, particularly for non-technical users. P02 argued that transparency was about clarifying what the system can and cannot do: \textit{``It's not clear to a user what the boundary is... the user doesn't know the limitations, the dos and the don'ts, the capabilities of that tool. That's the transparency of the capabilities that I want to resolve''} (P02). Similarly, P08 remarked:

\begin{quote}
    \textit{``Typically, you put out a product and you're like, here's all the things that it can do. Very rarely do you see, like, here's all the things that it can't do... I think especially now in this world of generative AI, it might actually be useful.''} (P08)
\end{quote}
Here, transparency served as a scaffolding mechanism to help users form realistic mental models and prevent over-reliance. Such transparency has less to do with making visible the internal workings of systems and more to do with communicating the scope and limits of system functionality to users.

\subsubsection{Transparency enables Internal Visibility}

In contrast, participants also viewed transparency primarily as a tool for developers rather than end users. For example, P03 suggested that multi-agent AI is still ``too early'' for broad user-facing transparency, instead arguing that it should initially be geared towards developers who need internal visibility for debugging: \textit{``While you must make things explicit, not just transparent, but also for debug ability and for developers, you want them to see the inner workings rather than hide them''} (P03).

Similarly, P05 highlighted that developers require detailed system visibility to isolate bugs, whereas end users benefit from higher-level summaries that build trust. P10 expanded this perspective, advocating for audit logs and system traces that would allow both developers and users to track agent communication step-by-step. Such mechanisms enable observability and foster iterative system refinement. To this point, P13 added:

\begin{quote}
    \textit{``From a developer standpoint, creating these agents transparency and observability means being able to see exactly what an agent is actually working on, and what is its context. I want to know at every moment, what does this agent know that is able to make that decision? Because that's where some corruption could happen.''}  (P13)  
\end{quote}
 These accounts suggest that transparency also functions as a diagnostic tool in development contexts.

\subsection{The ``How'' of Transparency: Mechanisms and Mediation}

\subsubsection{Transparency via Visual/Voice Communication of Agent Behavior}
A distinct perception linked transparency to UI design. P04 and P10 described the importance of visual interfaces that reveal how agents communicate and collaborate with each other. For instance, P04 envisioned UI metaphors such as group chats or process flow diagrams to make agent interactions comprehensible:

\begin{quote}
    \textit{``If the way they're [multiple agents] talking back and forth is a group chat, great. The UI is going to be a group chat, and you should be able to scroll through it and see all the messages. If it's more of like a business process flow... I would expect some directed graph visualization.''} (P04)    
\end{quote}
There are different mediums or modalities that were highlighted by these early adopters as a way to communicating the agent behavior other than chat interface. Participant P06 argued \textit{``Because agents talk to each other, doesn't mean that the interface is a chat... not everything is conversational. We should be agnostic... Let me talk to the agents of voice, there could be a channel sign.''} On the other hand, P09 compared the transparent articulation of the communication between agents visually like \textit{``you have UML type of diagrams, where it shows the relationships and the dependencies.''} 

Such visualizations were positioned as mechanisms to enable the verification and understanding of multi-agent orchestration. Situating transparency in UI affordances emphasizes that system comprehension relies not only on technical details but also on clear, intuitive, and understandable visualizations of agent-to-agent interactions.

\subsubsection{Transparency via Means to Critically Evaluate AI Outputs}

Finally, transparency was described as enabling the critical evaluation of AI outputs. P13 conceptualized the implementation of transparency as a form of critical evaluation, relying on metrics, fact checkers, or bias-detection tools to validate system outputs. P13 argued for ``lenses'' that could make biases visible: \textit{``I wish I could have telemetry so that I can showcase... how biased the content was. Just for example, in a non-fiction generator, is this bias towards only mentioning stories of men?''} (P13). By providing quantitative metrics and diagnostic means, systems empower users to independently verify generated content. Transparency here is instrumental in supporting both expert and lay evaluations of output quality. Adding to this point, P13 expanded the thought saying:

\begin{quote}
    \textit{``More like evaluated and validated by the experts in a way... potentially have the lenses that [experts] might have as a complementary tool to evaluate things as a post process. Basically, it's kind of like the word thing that you write something and it evaluates your tone, but instead it evaluates bias, intent, things you might not be aware of.''} (P13)
\end{quote}
These perspectives reveal that transparency in multi-agent LLM systems is a multi-faceted construct. Early adopters viewed it in various ways as open-source reproducibility, boundary definition, debugging support, UI visualization, trust-building, regulatory compliance, reactive traceability, and metric-based evaluation. The diversity of these framings suggests that transparency cannot be addressed using a single mechanism. Instead, it must be designed as a layered practice that aligns with developer needs, user comprehension, and institutional accountability.

\subsubsection{Transparency via Reactive and Event-Driven Traceability}

While many participants advocated for proactive transparency, others viewed it as being inherently reactive. P11 described transparency as often reactionary in nature, emerging primarily in response to  unexpected behaviors: \textit{``Developers usually start thinking about the transparency when things don't go as expected... usually people don't care until it is relevant for them''} (P11). Expanding on this, P11 adds:

\begin{quote}
    \textit{``The most basic stuff is like, what information did they fetch and why? And then, how did they pick the tool and why? And then, in the context of multi agents, the interactions between the agents is really important to figure out why something happened, because it's basically event-driven orchestration. So you need to understand, like an audit log of what happened.''} (P11)    
\end{quote}
This reactive orientation positioned audit logs as critical for reconstructing why unexpected behaviors occurred. P10 echoed this view, describing how transparency mechanisms such as step-by-step traces allow developers and users to \textit{``jump into different areas of the conversation and pinpoint areas.''} In this framing, transparency is less about continuous visibility and more about retrospective accountability.

\subsection{The ``Why'' of Transparency: Purposes}

\subsubsection{Transparency is needed for Open-Source Reproducibility}

For some participants, transparency was closely linked to openness and reproducibility. P01, for example, emphasized that transparency practices should make the system visible through openly shared models, code, and reproducibility packages. This approach was not only intended to enhance accountability but also to allow others to independently replicate findings:

\begin{quote}
    \textit{``Open-source models, open-source code that is exactly the architecture of what I have here. And some reports explaining the robustness of your models, maybe on the industry standard benchmarks and reproducibility package... Here are all the keys, all the tools, just insert your API key here.''} (P01)    
\end{quote}
Similarly, P10 commented that \textit{``...there's a surprising amount of really cool stuff, specifically with like the Llama style models and Hugging Face and everything else in terms of model comparisons and benchmarking''} as forms of community-driven reproducibility. By offering reproducibility packages, developers allow others to recreate results, verify findings, and build trust in the system's reliability. In this sense, transparency becomes a mechanism for accountability within the research and development community rather than for end-users alone.

\subsubsection{Transparency is needed for Auditability}
Participants working in compliance-heavy contexts described transparency as being essential for auditing. P07 emphasized the importance of the explainability of agent actions and traceability of data use, particularly in sensitive domains where violations could occur: \textit{``Transparency is more about how did you come up to generate that answer? If you were transparent on how an answer came to be, you can actually build the chain.''} (P07). Similarly, P09 highlighted the roles of model cards, privacy controls, and licensing verification by commenting:

\begin{quote}
    \textit{``From transparency practices... projects like Hugging Face, and even in AI studio, for example, many AI model providers do this. They provide a model card typically that model card tells you, here's the data that was used to train this model, here's the use cases that we know it works well in, here's the use cases where it doesn't... and here's potential biases that the model may have that you need to be aware of. So you're just very transparent again, about how this model was built, what it's good for, what it's not.''} (P09)    
\end{quote}
These mechanisms ensure that users understand the training data, potential biases, and data retention policies. By linking transparency to governance, participants framed it as a safeguard against misuse and a requirement for responsible AI deployment.

\subsubsection{Transparency is needed for Trust-Building}

Several participants perceived transparency as a trust-building mechanism, not just an epistemic tool. P05 argued that users benefit more from high-level summaries of model purpose and outcomes than from technical details: \textit{``From a developer perspective... any system you have a bug, you want to isolate a bug... I think for the user wise, of course, you need to help them to build up the trust of your system so you have... a clear way to explain how system works and what's the pros and cons.''}

P06 took this further by stressing that proof of functionality is paramount: \textit{``I think transparency is paramount to creating trust... they have to see what the system's doing... I think the first thing they would say is prove it.''} Similarly P01 highlighted the freedom of choice given to the users that builds the trust through higher agency:

\begin{quote}
    \textit{``At some point during the testing, we returned replies saying things like `Best models to estimate their confidence in reply' and this gave users some perspective to whether they could trust it or not that much. We thought about giving people more control over like `Hey only show me replies with more than something percent of confidence of reply'.''} (P01)    
\end{quote}
Trust-building perspectives highlight transparency as a pragmatic necessity. It assures users that the systems work reliably in practice, even when their internal mechanisms remain opaque. P08 added an ethical dimension to this perspective, suggesting that transparency should cover the impacts on stakeholders, fairness, and accessibility. For these participants, trust depends not only on functionality but also on the ethical framing of system development and use.

\section{Discussion: Towards a Multi-Dimensional Transparency Framework}

\begin{figure}
    \centering
    \includegraphics[width=0.5\linewidth, alt={Venn diagram with three circles showing dimensions of transparency in multi-agent LLM systems: Developer-focused (open-source models and codes, audit logs, reproducibility), User-focused (system boundaries, trust-building mechanisms, communication interfaces), and Governance-focused (explainable actions, traceability, model cards, licensing, consent controls, ethics). The overlap highlights Reactive and Proactive transparency.}]{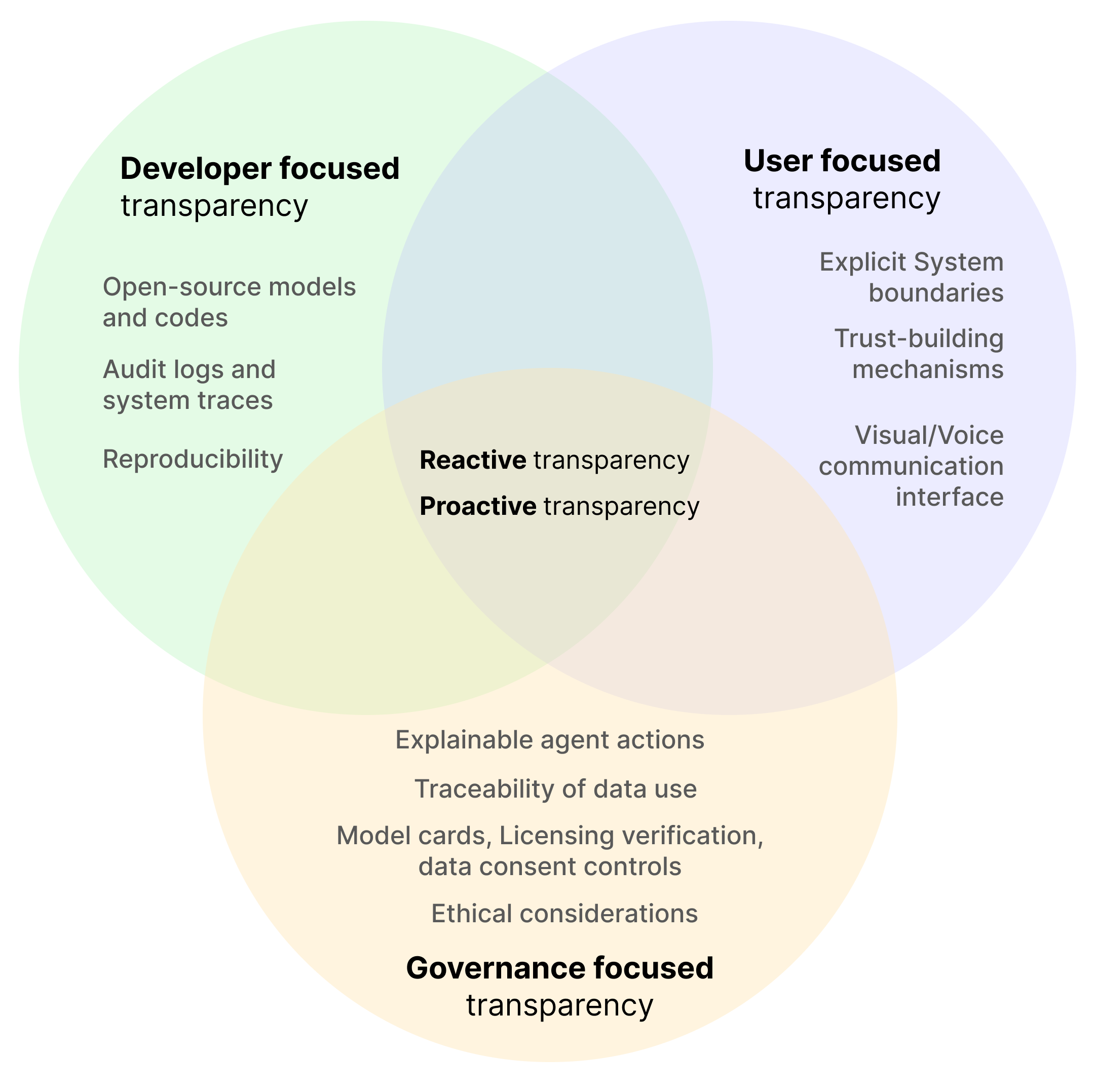}
    \caption{Multi-Dimensional Transparency Framework: Venn diagram illustrating three dimensions of transparency in multi-agent LLM systems: developer-focused, user-focused, and governance-focused. Their intersections highlight shared practices of reactive and proactive transparency.}
    \label{fig:transparency_framework}
\end{figure}

Our findings begin to scratch the surface of agentic transparency, demonstrating how early adopters articulate transparency in multi-agent LLM-based systems, which is not a singular or uniform concept. They framed transparency in terms of \textit{what} it entails, \textit{how} it is enacted, and \textit{why} it matters. Building on these accounts, our discussion advances a higher-level synthesis that extends beyond these diverse and sometimes-conflicting interpretations. Specifically, we propose a multidimensional framework (as shown in the figure ~\ref{fig:transparency_framework}) that positions transparency in terms of \textit{who} it is for and \textit{when} it becomes salient. We situate the \textbf{``who''} across three interrelated but distinct dimensions: \textbf{developer-focused transparency, user-focused transparency,} and \textbf{governance-focused transparency}. Together, these dimensions highlight the multifaceted nature of transparency as both a technical affordance and a socio-ethical construct. We conceptualize the \textbf{``when''} through the distinction between \textbf{reactive and proactive transparency}, which shows how transparency is contextually situated within adoption trajectories.

\subsection{Developer-Focused Transparency: Observability, Debugging, and Technical Reproducibility}
A central thread across participant accounts emphasized the role of transparency in enabling developers to understand, observe, and debug multi-agent systems. Participants such as P03, P05, and P10 stressed that transparency, at the current stage of system maturity, was ``premature'' for end users but essential for those building and maintaining these tools. From this perspective, transparency serves as an observability infrastructure, akin to debugging tools in conventional software engineering.

Mechanisms proposed to achieve this form of transparency included \textit{open-source models and codes} (P01) and \textit{audit logs and system traces} (P10) that expose agent reasoning and decision paths. These affordances enable developers to detect errors, isolate bugs, and refine agent orchestration. As P03 emphasized, developers ``want to see the inner workings rather than hide them,'' underscoring that transparency for some is less about simplification and more about deep system exposure.

Importantly, participants also linked developer-focused transparency to \textit{reproducibility}. P01 argued that open-source code and reproducibility packages enable others to replicate results, thus ensuring accountability and accelerating innovation. In this sense, transparency is positioned as both a technical necessity and a cultural practice rooted in open science.

Collectively, these accounts underscore that developer-focused transparency prioritizes depth over accessibility. It is designed for technically skilled actors who require full system visibility to ensure reliability, improve performance, and foster reproducibility of the results.

\subsection{User-Focused Transparency: Building Trust, Communicating Boundaries, and Designing Interfaces}

Another cluster of perspectives framed transparency as a means of enabling users, particularly non-technical users, to understand and trust multi-agent systems. Unlike developer-oriented transparency, which emphasizes detailed internal visibility, user-focused transparency aims to abstract complexity into interpretable and meaningful forms.

Three mechanisms stand out in this category. First, participants emphasized the importance of making \textit{system boundaries explicit}. P02 argued that users often lacked clarity on what the system could and could not do, which risked over-reliance or misuse. By explicitly communicating ``the dos and don'ts'' of system capabilities, transparency can scaffold realistic mental models and mitigate misaligned expectations.
Second, participants positioned transparency as a \textit{trust-building mechanism}. P05 and P06 described the value of high-level summaries of system purpose and outcomes, and the need to demonstrate functionality through proof-of-performance. As P06 articulated, users would often demand: ``prove it!'' This highlights that transparency is not merely about exposure but about providing assurances of system reliability in ways that resonate with users' concerns. Third, participants emphasized the \textit{role of interfaces} in transparency. P04 and P10 described how visualizing agent communication, whether in chat-like interfaces or process-flow diagrams, could help users make sense of otherwise opaque multi-agent coordination. These visual representations can reveal information flow, agent roles, and intermediate decisions in ways that are accessible without requiring technical knowledge or expertise.

Taken together, these findings suggest that user-focused transparency balances explainability and usability. It must abstract complexity without oversimplifying, ensuring that users can build trust, understand boundaries, and engage with outputs in an informed and empowered manner.

\subsection{Governance-Focused Transparency: Compliance, Ethics, and Accountability}

A third dimension situates transparency as a requirement for governance, regulation, and ethical accountability. Unlike developer- and user-focused transparency that primarily serve immediate operational needs, governance-focused transparency addresses broader societal and institutional imperatives.

Participants highlighted mechanisms that aligned with responsible AI practices. P07 described the need for \textit{explainable agent actions} and \textit{traceability of data use}, particularly in sensitive or regulated domains. In this sense, transparency enables organizations to detect potential data leaks, monitor compliance with privacy requirements, and justify decisions to external stakeholders. Similarly, P09 emphasized the \textit{roles of model cards, licensing verification, and data consent controls} as tools for ensuring institutional transparency. These mechanisms act as documentation and accountability infrastructures, making the provenance of training data, known biases, and permissible uses explicit. \textit{Ethical considerations} also shaped governance-focused transparency. P08 called for transparency to begin ``from the start of AI development,'' including articulating the system's social impact, fairness, and accessibility. Here, transparency was less about technical processes and more about ethical disclosure, making explicit the value-laden decisions embedded in design.

This governance-oriented framing positions transparency as an enabler of compliance, accountability, and legitimacy. This ensures that organizations can demonstrate due diligence, meet regulatory standards, and address societal expectations for responsible AI.

\subsection{Reactive versus Proactive Transparency}

An important distinction emerged between proactive and reactive transparency across these dimensions. Proactive approaches involve building transparency mechanisms into the system from the outset through \textit{open-source practices, user-facing explanations, or model cards}. In contrast, reactive approaches are \textit{triggered by failures, errors, or unexpected behaviors}. As P11 observed, transparency often becomes a priority ``when things don't go as expected.''

A few participants revealed that these two modes of transparency are not simply technical design choices but are bound up with adoption dynamics. For instance, P06 highlighted a perceived ``catch-22'': while transparency is crucial for long-term trust and integration into business operations, it is often not the central concern during early adoption. According to P06, prospective adopters may first need to be convinced that such systems ``actually work'' before transparency mechanisms become a requirement. In this framing, transparency shifts from being a proactive condition for uptake to a reactive demand that emerges once systems prove their utility and organizations begin relying on them.

This distinction has practical implications. Proactive transparency fosters trust and accountability before problems occur; however, it can be resource-intensive and difficult to prioritize during early development, as users are still uncertain about the system's value. Reactive transparency, while pragmatic, risks limiting transparency to damage control or post hoc justification, undermining broader commitments to openness and accountability. Therefore, an effective framework for multi-agent systems may need to integrate both, ensuring baseline proactive mechanisms while enabling reactive investigation when failures occur or uncertainties surface.

\subsection{Implications for HCI, AI Design and Research}

The synthesis of these perspectives starts to point towards an integrated view of transparency in multi-agent LLM systems. This framework positions transparency as:

\begin{enumerate}
    \item \textbf{Developer-Focused Transparency:} providing deep observability, debugging, and reproducibility tools for system builders.
    \item \textbf{User-Focused Transparency:} designing explanations, boundaries, and interfaces that foster understanding and trust among non-technical users.
    \item \textbf{Governance-Focused Transparency:} embedding accountability practices that satisfy regulatory, ethical, and institutional requirements.
\end{enumerate}

This framework emphasizes that transparency is not a singular technical feature but rather a socio-technical construct. Each dimension addresses different stakeholders, goals, and contexts of use, expanding beyond the primarily user-centered emphasis in XAI research \cite{ehsan_human-centered_2022, adadi_peeking_2018}. Importantly, these dimensions may complement or conflict with each other. For example, developer-focused transparency may expose details that overwhelm or confuse users, whereas user-focused abstractions may obscure information needed for compliance audits. Thus, designing for transparency in multi-agent systems requires careful negotiation across these dimensions, balancing depth, abstraction, and accountability as layers which was highlighted in Bansal et al. \cite{bansal_challenges_2024}.

From an HCI perspective, this framework extend prior work that framed transparency as a mechanism for explanation and trust calibration \cite{wang_are_2021, morrison_impact_2024}. We argue that transparency should instead be approached as a design problem of \textit{identification and translation for action} rather than mitigation. System designers must ask: \textit{who} is this system being made transparent to, and \textit{what forms of information are most actionable for them}? This extends beyond technical instrumentation. This highlights the need for selectivity and directedness in transparency practices, tailoring mechanisms to specific audiences. Visualization, trust-building mechanism, boundary-setting, and interaction design emerge as critical tools for shaping user-focused transparency, while debugging, auditing, traceability, reproducibility, documentation, and open-source practices remain central to developer and governance concerns. 

Moreover, the findings expand on XAI work that primarily emphasizes model-level interpretability and explanation \cite{wang_are_2021, adadi_peeking_2018} by surfacing transparency as a relational construct in multi-agent contexts. Trust, accountability, and reproducibility are not properties of the system alone but are co-constructed between system builders, users, and institutions \cite{wang_investigating_2024, zhang_i_2024}. This complements existing research on standardized documentation tools such as model cards, datasheets, and fact sheets \cite{liao_designerly_2023}, while adding evidence from early adopters that reproducibility packages and open source practices are essential for accountability in multi-agent settings. 

Finally, the study validates concerns raised in prior work on the unique transparency challenges of multi-agent generative systems \cite{bansal_challenges_2024}. Our findings argues that distributed orchestration, inter-agent dependencies, and emergent behaviors deepen the ``black box'' effect and demand mechanisms for event-driven traceability, retrospective audit logs, and orchestration-level visualization. At the same time, they expand the conversation by illustrating how early adopters conceptualize transparency not only as a tool for comprehension but also as a diagnostic mechanism for debugging, a trust scaffold for users, and an accountability practice for governance. In this sense, transparency in multi-agent LLM systems is best understood as a layered practice that is situated within broader interaction and organizational contexts, aligning with the expectations and capacities of their intended audiences.

\section{Future Work}
Future studies should examine how transparency needs shift across the lifecycle of multi-agent AI systems. In early development stages, developers emphasize observability, debugging, and reproducibility, whereas later stages of deployment introduce new demands for user comprehension and regulatory compliance. Longitudinal studies can track the evolution of transparency practices as systems mature, from experimental prototypes to workplace tools and consumer applications, revealing how priorities, mechanisms, and trade-offs change over time. 
While our study focused on early adopters in a large technology organization, future research should explore transparency perceptions in high-stakes sectors such as healthcare, finance, and law. In these domains, multi-agent systems are beginning to emerge; however, errors, opacity, and lack of accountability may have additional and significant consequences. Investigating how professionals and end users in these contexts interpret transparency would reveal domain-specific requirements and ethical challenges and could inform governance strategies tailored to sensitive applications.
Finally, future research should move beyond descriptive accounts toward the participatory design of transparency mechanisms. Co-design approaches involving developers, end users, and regulators can surface shared concerns while accommodating divergent needs. Prototyping transparency tools, such as visualizations of agent interactions, layered explanation interfaces, or audit infrastructures, offer an opportunity to evaluate how different mechanisms foster trust, accountability, and usability. Such participatory methods would not only produce actionable design strategies but also ensure that transparency practices are grounded in stakeholder experiences.

\section{Conclusion and Limitations}
In this paper, we examined how early adopters of multi-agent LLM-based systems perceive and practice transparency. Through semi-structured interviews with 13 participants in a [large technology organization], we surfaced diverse and sometimes conflicting interpretations of transparency, ranging from reproducibility and debugging to trust-building, governance, and ethical accountability. These findings demonstrate that transparency in multi-agent systems is not a singular concept but a layered and situated practice shaped by context, stakeholder roles, and system maturity.

Our analysis contributes to the literature in three ways. Empirically, we provide one of the first qualitative accounts of transparency in the emerging domain of multi-agent AI. Conceptually, we advance the notion of transparency as a situated practice, showing how developer-, user-, and governance-focused perspectives coexist. We discuss the implications for HCI and AI design, emphasizing the need for transparency mechanisms that adapt to stakeholder roles, balance proactive and reactive strategies, and evolve alongside system maturity.

As foundational research, this study has some limitations. First, our participant sample was relatively small (n=13) and drawn from a single large technology organization, which may not fully capture the diversity of perceptions about transparency. Early adopters in this setting may have access to resources, infrastructure, and expertise that differ from those in smaller companies, academia, or independent developer communities. Second, while our interpretivist approach allowed participants' own framings of transparency to emerge, it also meant that the findings reflected perceptions rather than objective measures of system transparency. Third, because multi-agent LLM systems remain in the early stages of development, participants' accounts may shift as these systems mature, scale, and become subject to regulatory oversight.

By centering on early adopters, we highlight how transparency is negotiated in real-world contexts before norms and standards solidify. Although early adopters are not transparency experts, their perspectives matter; they reveal practical challenges, emergent expectations, and design opportunities that will shape how transparency becomes embedded in the everyday use of multi-agent systems. Looking forward, our study opens several avenues for future research. HCI/AI researchers should investigate how transparency needs shift across the life cycle of multi-agent LLM-based AI systems, from early stage development focused on debugging and reproducibility to mature deployments that demand user-facing explanations and regulatory accountability. Future studies should also explore transparency in high-stakes sectors such as healthcare, finance, and law, where system opacity carries heightened ethical and social risks. Finally, participatory approaches to co-designing transparency tools with developers, end-users, and regulators can help ensure that mechanisms such as visualization interfaces, layered explanations, and audit infrastructures align with the needs of diverse stakeholders.

As multi-agent LLM-based AI systems continue to expand into professional and consumer domains, the question is not only how to make them technically transparent but also how to design transparency as a socio-technical construct that serves diverse stakeholders. Addressing this challenge requires moving beyond a single definition of transparency toward a pluralistic approach that recognizes transparency as relational, negotiated, and situated in practice.

\begin{acks}
to be added
\end{acks}

\bibliographystyle{ACM-Reference-Format}
\bibliography{AI_references}

\appendix

\begin{table*}
\caption{Transparency perceptions and mechanisms across participants (part 1 of 2)}
\label{tab:transparency_findings1}
\begin{tabular}{p{0.25\linewidth} p{0.35\linewidth} p{0.32\linewidth}}
\hline
\textbf{Transparency perception} & \textbf{Mechanisms} & \textbf{P's Quote}\\ \midrule
\textbf{Open-source and reproducibility} & \textit{Open-source models and code:} Using these models and providing open-source code to enhance transparency. This makes the architecture, models, and processes visible to others, allowing users and developers to see exactly how the system operates in practice. 

\textit{Reproducibility packages:} By providing these packages, users can recreate the same results using the provided code and tools. & P01: \textit{``Open-source models, Open-source code that is exactly the architecture of what I have here. And some reports explaining the robustness of your models, maybe on the industry standard benchmarks and reproducibility package...''}\\

\hline

\textbf{AI's capability boundaries and limitations} & \textit{Explicit Boundaries:} The need for making the capability, limitation, and boundaries of AI tools explicit. This is mostly for making non-technical users understand what the tool can and cannot do. & P02: \textit{``multi-agent tool is very small scope but for all the generative AI tools, as it's not clear to a user what the boundary is. So in the end the user doesn't know the limitations, the dos and the don'ts, the capabilities of that tools. That 's the transparency of the capabilities that I want to resolve.''}\\

\hline 

\textbf{Transparency as debugging} & \textit{Developer-oriented and internal visibility for debugging}: considers transparency to be premature at the current development stage. If addressed, it should gear towards making inner workings visible for developers; similar to no-code/low-code debugging needs. & P03: \textit{``While you must make things explicit... for debug ability and for developers, you want them to see the inner workings rather than hide them.''}\\

\hline

\textbf{UI for agent communication} & \textit{Graphical representation of agent interaction:} visualize agent interactions and message flows (e.g., group chat views or directed graphs) to surface how agents collaborate. & P04: \textit{``If the way they're talking back and forth is a group chat... The UI is going to be a group chat... If it's more of like a business process flow... I would expect some directed graph visualization.''}\\

\hline

\textbf{Trust-building} & \textit{High-level summaries and model purpose to build trust:} For non-technical users, participants suggested using mechanisms that provide high-level summaries of the model's purpose and the outcomes it is designed to achieve. Instead of showing the intricate details of agent orchestration, these summaries focus on the system 's goals and overall impact. & P05: \textit{``From a developer perspective, I think that I would say, you want to see it won't have a good way to depart the system. Any system you have a bug, you want to isolate a bug and you can resolve it and then make the tool better.''}\\

\hline

\textbf{Trust-building} & \textit{Proof of functionality:} It is important to first demonstrate the effectiveness of the tool as a way to build trust. \textit{Visibility to agent's actions:} perceives transparency in terms of understanding the actions performed by agents. & P06: \textit{``I think transparency is paramount to creating trust... they have to see what the system's doing... I think the first thing they would say is prove it.''}\\

\hline

\end{tabular}
\end{table*}

\begin{table*}
\caption{Transparency perceptions and mechanisms across participants (part 2 of 2)}
\label{tab:transparency_findings2}
\begin{tabular}{p{0.25\linewidth} p{0.35\linewidth} p{0.32\linewidth}}
\hline

\textbf{Trust-building} & \textit{Summaries for ethical transparency:} Think of transparency from the start of the AI development and cover multiple dimensions like impact, affected stakeholders, data collection, training, and intended purpose (including feasibility, fairness, accessibility). & P08: \textit{``...make it more clear how the AI is... what is the impact of AI?... who does it affect?... how is the data collected? How is the model trained? what is the purpose of the model at the end of the day.''}\\

\hline
\textbf{Regulatory compliance and auditing-oriented} & \textit{Explainable agent actions:} It is important to have the agents explain how they arrived at a decision, including data sources used and the processing steps taken. This ensures that users can trace the logic behind the decision, allowing them to understand how conclusions are derived and assess the reliability of the result. & P07: \textit{``The transparency is more about how did you come up to generate that answer? ... If you were transparent on how an answer came to be, you can actually build the chain.''}\\

\hline

\textbf{Regulatory compliance and auditing-oriented} & \textit{Model cards, Privacy, licensing:} This includes having model cards listing training data, intended use, failure cases, and potential biases; user control over data retention/consent; early licensing verification to avoid misuse and bias risks. & P09: \textit{``Projects like hugging face provide a model card. Here's the data that was used, the use cases that we know, where it doesn't work, potential biases. Being transparent about how this model was built.''}\\

\hline

\textbf{Reactive and event-driven traceability} & \textit{Trails of agent's action and decision:} focuses on understanding the decision-making process of conversational agents, emphasizing traceability and accountability. There is a need for detailed information about what actions were taken by agents, why those actions were chosen, and how specific tools were selected. & P11: \textit{``...in the context of multi agents, the interactions between the agents, just the order and the interactions between the agents is really important to figure out why something happened... So you need to understand how, like an audit log of what happened.''}\\

\hline

\textbf{Reactive and event-driven traceability} & \textit{Audit logs and system traces:} maintaining the audit logs and records of all actions taken by the agents to ensure traceability. This would allow developers to pinpoint when and where the errors occurred. & P10: \textit{``if you ever use these systems there, they talk a lot and it's so much information that is very difficult to understand what is going on or to keep track of it. So how can you build a system where a you can jump into different areas of the conversation and pinpoint areas?''}\\

\hline

\textbf{Critical evaluation mechanism by the user} & \textit{Metric-based and verification oriented transparency:} perceives transparency as an ability to evaluate and verify the AI's output using specific metrics and tools. Importance of providing users with lenses - such as bias-checker, fact-checker, and other validation metrics - to assess the generated content effectively. & P13: \textit{``I don't see anything wrong with that, as long as you have some sort of metric of things that you might help... have metrics to show what biases might be present. Whatever is going to make up things also, like fact checker, just having these lenses that will help you actually verify those things might be useful to have a transparency.''}\\

\hline
\end{tabular}
\end{table*}

\end{document}